# Atomically resolved structural determination of graphene and its point defects via extrapolation assisted phase retrieval

Tatiana Latychevskaia and Hans-Werner Fink

Physics Department, University of Zurich

Winterthurerstrasse 190, 8057 Zurich, Switzerland

**Abstract**

Previously reported crystalline structures obtained by an iterative phase retrieval reconstruction of their diffraction patterns seem to be free from displaying any irregularities or defects in the lattice, which appears to be unrealistic. We demonstrate here that the structure of a nanocrystal including its atomic defects can unambiguously be recovered from its diffraction pattern alone by applying a direct phase retrieval procedure not relying on prior information of the object shape. Individual point defects in the atomic lattice are clearly apparent. Conventional phase retrieval routines assume isotropic scattering. We show that when dealing with electrons, the quantitatively correct transmission function of the sample cannot be retrieved due to anisotropic, strong forward scattering specific to electrons. We summarize the conditions for this phase retrieval method and show that the diffraction pattern can be extrapolated beyond the original record to even reveal formerly not visible Bragg peaks. Such extrapolated wave field pattern leads to enhanced spatial resolution in the reconstruction.

**Main text**

The study of nanocrystal structures at atomic resolution is an important topic in nanotechnology, solid state physics and especially in biology, where preparing a large perfect crystal is often a challenge and the synthesis of nanocrystals is preferred[1]. It has recently been demonstrated that the structure of a nanocrystal can directly be obtained by coherent diffraction imaging[2] from an electron or X-ray diffraction pattern by applying phase retrieval algorithms[3-7]. The diffraction pattern of a crystalline structure typically consists of distinct Bragg peaks, whereby each peak is convoluted with the Fourier transform of the crystal shape (shape-transform)[8-9]. In the experiments demonstrated so far[3-7], a regular crystalline structure at sub-nanometer resolution could be retrieved, but the individual atoms remained unresolved and therefore no atomic defects were revealed. The structure retrieval in the reported experiments require the input of an initial low-resolution image of the sample distribution typically provided by other techniques, as for example by transmission electron microscopy (TEM) imaging[3-4], scanning electron microscopy (SEM)[5], high-resolution transmission electron microscopy (HRTEM) imaging[6] or holography[7].

Here, we address the problem of uniqueness of the crystalline structures obtained by phase retrieval from diffraction patterns. It is obvious that the Fourier transform of a diffraction pattern exhibiting distinct Bragg peaks will always result in some periodic structure that will remain being a periodic structure under further phase retrieval. This raises the question whether the previously reported reconstructed periodic structures actually reflect the true distribution of atoms in the crystal. The apparent perfect periodicity free from displaying any irregularities or defects in the lattice in most published reconstructions hints to this as being an important issue. To answer this question we simulated a diffraction pattern of a crystal with atomic scale defects using realistic electron scattering amplitudes and setting the conditions for recovering the true crystalline structure together with its atomic scale defects.

The resolution of the reconstructed sample is given by the highest order scattering signal detected in the diffraction pattern at an angle $\vartheta_{max}$ which defines the numerical aperture of the setup and the wavelength $\lambda$: $\Delta_0 = \frac{\lambda}{2\sin\vartheta_{max}}$. Thus, to resolve individual atoms, the wavelength of the probing wave, the sample to detector distance and the detector size must be selected such that $\Delta_0$ is less than the interatomic distances. As a test sample we select a graphene patch with two defects: a divacancy and a trivacancy as shown in Fig. 1(a). The shortest distance between carbon atoms in graphene amounts to 1.42 Å and therefore the parameters of the simulations presented here are selected such that $\Delta_0 = 50$ pm.

A diffraction pattern can only unambiguously be reconstructed when the oversampling condition is being fulfilled[2]. This implies that the area occupied by the sample in the object domain must be enclosed in a known support of at least twice the size of the sample. Just the appearance of Bragg peaks alone, as they have already been observed in the early famous Laue type X-ray experiments or the Davisson–Germer type electron scattering experiments, leave local details of the sample for ever uncovered since the oversampling condition is not fulfilled here. The oversampling condition implies that when an experimental record is digitized with N × N pixels, the size of the reconstructed area is $\Delta_0 N \times \Delta_0 N$ and thus, the area occupied by the sample must be limited to $0.5\Delta_0 N \times 0.5\Delta_0 N$ in size at the largest. Experimentally, this condition is often fulfilled by either limiting the beam size [3] or by employing a finite aperture in the object plane[7]. In our simulation we assume that the graphene sample is mounted over an aperture that limits the size of the area exposed to the wave front to 6 nm in diameter, which provides an oversampling ratio of approximately 7.5. The positions of individual atoms are provided as their exact spatial coordinates (not pixels). For electron scattering the complex-valued amplitudes were constructed as the partial wave expansion[10]:

$$f(\vartheta) = \sum_{l=0}^{\infty}(2l+1)\frac{1}{k}e^{i\delta_l(k)}\sin\delta_l(k)P_l(\cos\vartheta), \qquad (1)$$

where $k$ is the wave number, $P_l(\cos\vartheta)$ are Legendre polynomials, $\vartheta$ is the scattering angle, $l$ is the angular momentum number for each partial wave ($l$=0 corresponds to isotropic s-waves, and so on), and $\delta_l(k)$ are the phase shifts. The complex-valued scattering amplitudes were calculated using phase shifts $\delta_l(k)$ provided by the NIST library[11] for high-energy (20 keV) electrons and by the van Hove phase shift package[12] for low-energy (300 eV) electrons whereby a graphene patch was created for being a realistic scattering object. The complex-valued waves scattered off each atom were superimposed in the far-field and the intensity of the total wave field provides the diffraction pattern.

Figure 1(b) shows such diffraction pattern of the graphene patch in $q$ coordinates, simulated for 300 eV electrons and sampled with 1000 × 1000 pixels. Bragg peaks up to the third order are observed. Figure 1(c) shows the distributions of the scattering amplitudes $|f(\vartheta)|^2$ for three types of the scattering processes: isotropic scattering (s-waves) and anisotropic scattering for high- and low-energy electrons. For electrons, the amplitude of the scattered wave has pronounced maxima in the direction of the incident wave. The higher the energy of the probing electrons the more pronounced is the effect of forward scattering as apparent from the scattering amplitudes calculated for 20 keV and 300 eV shown in Fig. 1(c). Figure 1(d) depicts the intensity profiles of the diffraction patterns calculated with these three types of the scattering amplitudes. For isotropic scattering, the intensity of the second order peaks is higher than that of the first order peaks. However, for diffraction of electrons, the effect of strong forward scattering leads to the fact that the intensity of the first order peaks is always higher than the intensity of the second order peaks. In a TEM diffraction experiment, the relative intensities of the first to the second order peaks allows to distinguish between single and bi-layer graphene: for bi-layer graphene the second order peaks exhibit a lower intensity compared to the first order[13-14].

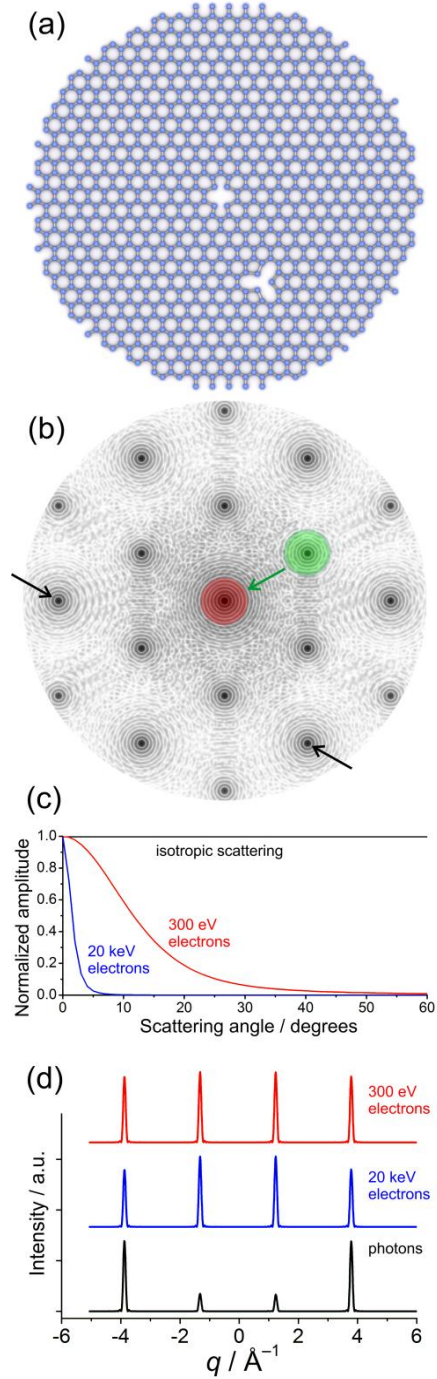

FIG. 1 Diffraction pattern of the graphene patch with two defects: a divacancy and a trivacancy. (a) Choice of the atomic arrangement. (b) Diffraction pattern simulated for electrons of 300 eV kinetic energy, shown in inverted intensity and logarithmic scale. For phase retrieval reconstruction, the central region of 110 pixels in diameter (indicated by a red circle) is replaced by the two-dimensional intensity distribution of one of the first order peaks (indicated by a green circle). (c) Normalized scattering amplitudes $|f(\vartheta)|^2$ shown for the three cases: isotropic scattering, 20 keV

and 300 eV electron energy scattering. (d) Intensity profiles along the line indicated by arrows in (b) shown for three cases: isotropic scattering, 20 keV and 300 eV energy electron scattering.

Each peak in a diffraction pattern of a crystalline nanostructure corresponds to the convolution of an ideal, delta-function like peak with the Fourier transform of the crystal shape[8]. With sufficiently fine sampling, the overall shape of the nanocrystal can even be reconstructed from the intensity distribution of just one of the peaks and its surrounding region[1,15-19]. In the simulations presented here, the central region of the diffraction pattern of 110 pixels in diameter, as indicated in Fig. 1(b) by a red circle, was assumed to be missing in order to mimic realistic experimental conditions where the central region is either blocked or overexposed. However, the central region in a diffraction pattern is required for stable convergence of the phase retrieval routine as it provides information about the low-resolution shape of the sample. This missing region was replaced with the two-dimensional intensity distribution of one of the first-order diffraction peaks, as illustrated in Fig. 1(b) by a green circle, whose intensity was scaled up by a factor of 20. The factor 20 was derived from the theoretical ratio of the intensities of the zero to first order diffraction peaks of graphene.

The reconstruction was done by applying the most popular hybrid-input-output phase retrieval algorithm[20] for 300 iterations with an initial random phase distribution and the feedback parameter $\beta$ = 0.9. All phase retrieval algorithms are based on forward and backward propagation of the scattered wave field between sample and detector plane. Within the approximation of isotropic scattering this boils down to just applying forward and backward Fourier transforms, respectively. In reality, only photon scattering is isotropic whereas electrons of any energy scatter with strongly enhanced amplitudes in forward direction. Thus, assuming isotropic scattering for electrons and applying simple Fourier transforms for wave propagation in phase retrieval routines is a rough approximation, which however provides meaningful reconstruction, as illustrated in Fig. 2. It is in fact a fortunate situation that the integrals governing coherent optics and being originally designed for isotropic light scattering are also applicable for anisotropic scattering processes. What is lacking, however, is the quantitative reconstruction of the object transmission function. The amplitude of the reconstructed complex-valued distribution at the sample plane exceeds unity since it is dominated by the strong signal from scattering atoms. However, this constitutes an unphysical condition for a realistic transmission function. Given that, it is worth to note that the correct transmission function, as for example the contour of the aperture can only be reconstructed when the scattering is assumed to be isotropic, compare Fig. 2(a) to Fig. 2(b) and (c). It has previously already been shown that the additional constraint of non-negative absorption facilitates the convergence of the algorithm [21]. Here, a modified constraint of a limited object scattering amplitude was applied by forcing the amplitude of the scattered wave below a threshold of 6.

The defects are revealed at the exact locations where they were originally positioned for the simulation, see Fig. 2. While the atomic positions were provided in spatial coordinates during the simulation, the reconstruction is digitally sampled and the position of each reconstructed atom is thus distributed over a few neighbouring pixels.

It should be noted that it is not possible to reconstruct the graphene structure when only the first six peaks are available in the diffraction pattern. This is due to the fact that a similar diffraction pattern exhibiting six fold symmetry corresponds to a trigonal lattice and a phase retrieval routine quickly converges and stagnates at such trigonal structure, see Fig. 2(c).

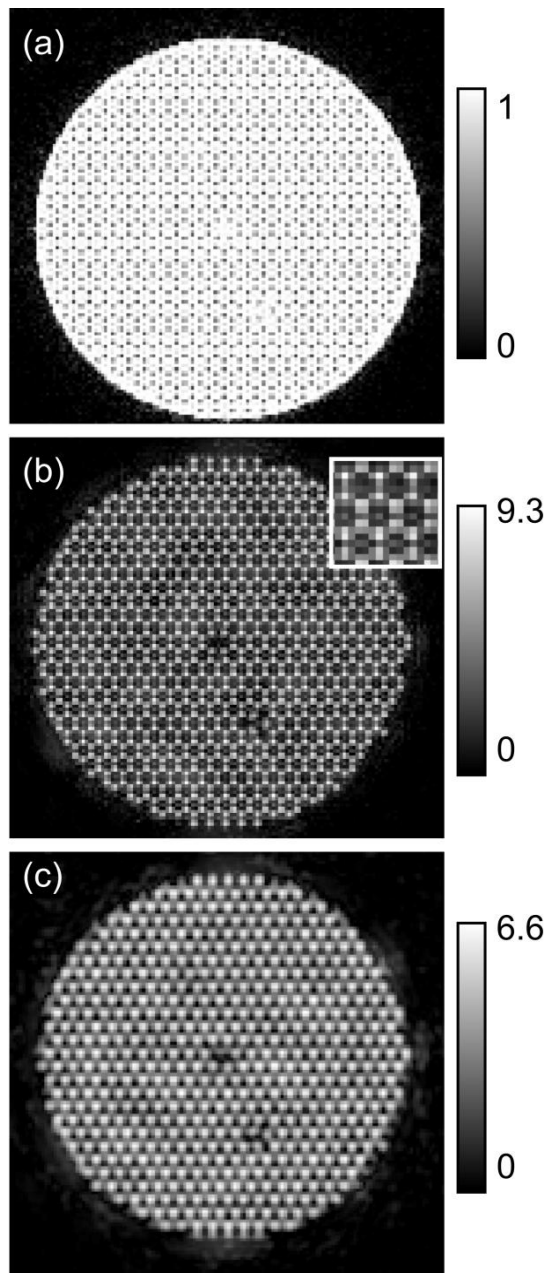

FIG. 2 Amplitude of the transmission function of the graphene patch with two defects: a divacancy and a trivacancy reconstructed by phase retrieval

from a diffraction pattern simulated with (a) isotropic s-wave scattering and (b) anisotropic scattering of 300 eV electrons. The inset shows a magnified fragment of the reconstruction. (c) Reconstruction for 300 eV electrons when only the six first order peaks are available in the diffraction pattern.

For realizing coherent diffraction in an experiment, the following requirements need to be fulfilled. (1) The beam must exhibit sufficient spatial and transversal coherence to exceed the sample size. This can be achieved by placing the sample onto a small aperture whose size may not exceed the coherence length of the beam. (2) The oversampling ratio must be more than 2. The higher the oversampling ratio the faster the convergence of the phase retrieval routine. (3) There should be sufficient signal in the diffraction pattern at $q$-numbers related to the required resolution[22]. For example, Zuo[3] has reported atomic reconstructions from an electron diffraction pattern recorded at a current density of $10^5$ e/(s·nm$^2$). The last requirement implies that the probing wave must have sufficient intensity and/or one must integrate over a sufficiently long acquisition time. In order to study the effect of different radiation doses and thus the related signal-to-noise ratio onto the reconstruction results, we simulated and reconstructed diffraction patterns for electrons of 300 eV energy at different total electron doses. We also added Gaussian distributed noise with the mean equal to the square root of the intensity at a pixel. Figure 3 shows the results. With a total electron dose of $10^{10}$ e/nm$^2$, the two defects: a divacancy and a trivacancy are clearly retrieved (Fig. 3(a)). With a total electron dose limited to just $10^6$ e/nm$^2$, the two defects will not be resolved anymore in the reconstruction (Fig. 3(e)). In the simulations we assumed that 1 electron scattering event results in 1 count per pixel, which in reality varies depending on the efficiency of the detecting system.

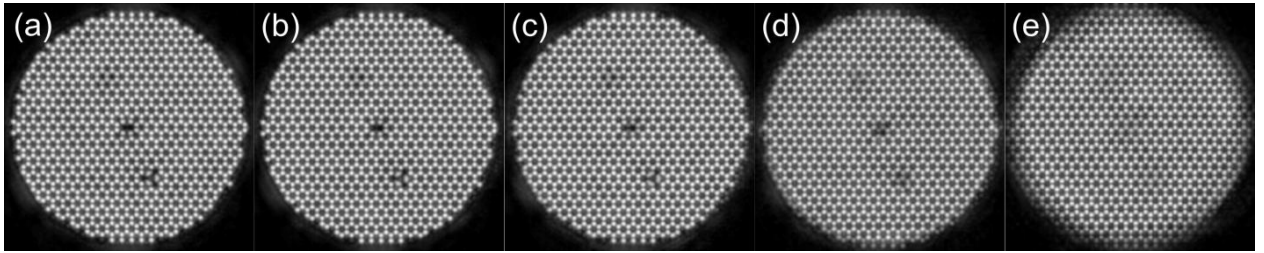

FIG. 3 Reconstructed simulated diffraction patterns of the graphene sample with two defects: a divacancy and a trivacancy, taken with 300 eV electrons at different total electron dose: (a) $10^{10}$ e/nm$^2$, (b) $10^9$ e/nm$^2$, (c) $10^8$ e/nm$^2$, (d) $10^7$ e/nm$^2$, (e) $10^6$ e/nm$^2$. Each reconstruction is the result of averaging of 10 successful reconstructions.

Recently, following the idea of obtaining super-resolution in an image[23-24], it has been demonstrated that a diffraction pattern of a continuous sample can be extrapolated beyond the experimentally detected area[25]. Here, we apply the same extrapolation method to a diffraction pattern of a crystalline sample with the results shown in Fig. 4. The details of the extrapolation procedure can be found elsewhere[25-26]. In brief, the complex-valued wavefront distribution reconstructed by a conventional phase retrieval algorithm is padded with random complex-valued numbers up to 2000 × 2000 pixels. The random padding in Fourier domain was updated after each iteration. The distribution in the central spot of the diffraction pattern was kept equal to the one recovered by the conventional phase retrieval algorithm. Otherwise, when the central spot is kept free from this constraint and updated after each iteration, the extrapolation fails. A constraint of limited amplitude of the scattered wave was applied in the object domain.

Figure 4(a) shows the originally available diffraction pattern and its extrapolated part. Newly revealed Bragg peaks appear in the extrapolated diffraction pattern, although some of them exhibit a double peak appearance due to unavoidable finite sampling by square pixels in the object domain.

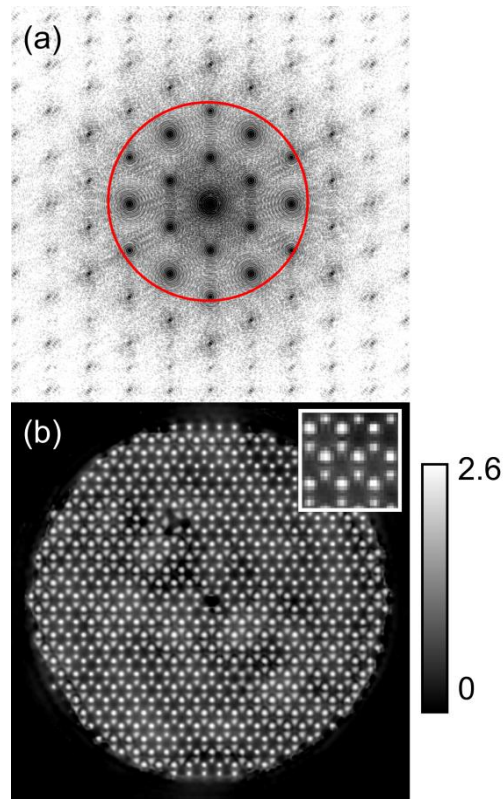

FIG. 4 Extrapolated diffraction pattern and its reconstruction. (a) The original diffraction pattern displayed inside the red circle has been extrapolated outwards. (b) Its reconstruction. The inset shows a magnified fragment of the reconstruction.

The padding in Fourier domain towards 2N × 2N pixels does not change the physical size of the reconstructed object area, but effects only its sampling to 2N × 2N pixels. As a result, the pixel size in the object domain decreases to $\Delta_0/2 = 25$ pm which allows a more precise localization of the atomic positions. Besides this effect, even more important is the effective increase of the numerical aperture as a result of the extrapolation of the diffraction pattern, which leads to an improved resolution in the reconstruction, as evident from the inset in Fig. 4(c).

We have demonstrated that a crystalline structure of nanometer dimension can be retrieved from its diffraction pattern alone without the need of additional low-resolution image information about the shape of the object. The atomic defects reconstructed in the recovered structure validate the non-ambiguity of the reconstruction. However, the transmission function of the sample imaged with electrons cannot be quantitatively retrieved from its diffraction pattern, because the phase retrieval routines are based on the assumption of isotropic scattering while electrons of any energy scatter with amplitudes exhibiting strong maxima in the direction of the incident wave. Furthermore, we have demonstrated that a diffraction pattern of a crystalline structure can numerically be post-extrapolated towards a larger numerical aperture which *a posteriori* increases the resolution of the retrieved nanostructure.

**Acknowledgements**

The authors are grateful to Prof. Michel A. van Hove of Hong Kong Baptist University for his assistance in calculating the scattering phase shifts by using his low-energy electron diffraction software package. The work presented here has financially been supported by the Swiss National Science Foundation (SNF) and the University of Zurich.

**References**

[1] H. N. Chapman, P. Fromme, A. Barty, T. A. White, R. A. Kirian, A. Aquila, M. S. Hunter, J. Schulz, D. P. DePonte, U. Weierstall, R. B. Doak, F. R. N. C. Maia, A. V. Martin, I. Schlichting, L. Lomb, N. Coppola, R. L. Shoeman, S. W. Epp, R. Hartmann, D. Rolles, A. Rudenko, L. Foucar, N. Kimmel, G. Weidenspointner, P. Holl, M. Liang, M. Barthelmess, C. Caleman, S. Boutet, M. J. Bogan, J. Krzywinski, C. Bostedt, S. Bajt, L. Gumprecht, B. Rudek, B. Erk, C. Schmidt, A. Homke, C. Reich, D. Pietschner, L. Struder, G. Hauser, H. Gorke, J. Ullrich, S. Herrmann, G. Schaller, F. Schopper, H. Soltau, K.-U. Kuhnel, M. Messerschmidt, J. D. Bozek, S. P. Hau-Riege, M. Frank, C. Y. Hampton, R. G. Sierra, D. Starodub, G. J. Williams, J. Hajdu, N. Timneanu, M. M. Seibert, J. Andreasson, A. Rocker, O. Jonsson, M. Svenda, S. Stern, K. Nass, R. Andritschke, C.-D. Schroter, F. Krasniqi, M. Bott, K. E. Schmidt, X. Wang, I. Grotjohann, J. M. Holton, T. R. M. Barends, R. Neutze, S. Marchesini, R. Fromme, S. Schorb, D. Rupp, M. Adolph, T. Gorkhover, I. Andersson, H. Hirsemann, G. Potdevin, H. Graafsma, B. Nilsson, and J. C. H. Spence, Nature **470** (7332), 73 (2011).

[2] J. W. Miao, P. Charalambous, J. Kirz, and D. Sayre, Nature **400** (6742), 342 (1999).

[3] J. M. Zuo, I. Vartanyants, M. Gao, R. Zhang, and L. A. Nagahara, Science **300** (5624), 1419 (2003).

[4] W. J. Huang, J. M. Zuo, B. Jiang, K. W. Kwon, and M. Shim, Nature Phys. **5** (2), 129 (2009).

[5] O. Kamimura, T. Dobashi, K. Kawahara, T. Abe, and K. Gohara, Ultramicroscopy **110** (2), 130 (2010).

[6] L. De Caro, E. Carlino, G. Caputo, P. D. Cozzoli, and C. Giannini, Nature Nanotech. **5** (5), 360 (2010).

[7] J.-N. Longchamp, T. Latychevskaia, C. Escher, and H.-W. Fink, Phys. Rev. Lett. **110** (25), 255501 (2013).

[8] M. v. Laue, Annalen der Physik **26** (5), 55 (1936).

9. J. C. H. Spence, R. A. Kirian, X. Y. Wang, U. Weierstall, K. E. Schmidt, T. White, A. Barty, H. N. Chapman, S. Marchesini, and J. Holton, Opt. Express **19** (4), 2866 (2011).
10. L. D. Landau and E. M. Lifshitz, *Quantum Mechanics: Non-Relativistic Theory (Course of Theoretical Physics)*, Third Edition ed. (Pergamon Press, Oxford, 1977).
11. NIST, *NIST electron elastic-scattering cross-section database*. (2000).
12. M. A. Van Hove, Van Hove phase shift calculation package.
13. J. C. Meyer, A. K. Geim, M. I. Katsnelson, K. S. Novoselov, D. Obergfell, S. Roth, C. Girit, and A. Zettl, Solid State Commun. **143** (1–2), 101 (2007).
14. R. Zan, Q. M. Ramasse, R. Jalil, and U. Bangert, in *Advances in Graphene Science*, edited by Mahmood Aliofkhazraei (InTech, 2013).
15. I. K. Robinson, I. A. Vartanyants, G. J. Williams, M. A. Pfeifer, and J. A. Pitney, Phys. Rev. Lett. **87** (19), 195505 (2001).
16. G. J. Williams, M. A. Pfeifer, I. A. Vartanyants, and I. K. Robinson, Phys. Rev. Lett. **90** (17), 175501 (2003).
17. I. Robinson and R. Harder, Nature Mater. **8** (4), 291 (2009).
18. R. Harder, M. Liang, Y. Sun, Y. Xia, and I. K. Robinson, New J. Phys. **12** (2010).
19. M. C. Newton, S. J. Leake, R. Harder, and I. K. Robinson, Nature Mater. **9** (2), 120 (2010).
20. J. R. Fienup, Appl. Optics **21** (15), 2758 (1982).
21. T. Latychevskaia and H.-W. Fink, Phys. Rev. Lett. **98** (23), 233901 (2007).
22. R. Neutze, R. Wouts, D. van der Spoel, E. Weckert, and J. Hajdu, Nature **406** (6797), 752 (2000).
23. R. W. Gerchberg, Opt. Acta **21** (9), 709 (1974).
24. A. Papoulis, IEEE Transactions on Circuits and Systems **22** (9), 735 (1975).
25. T. Latychevskaia and H.-W. Fink, Appl. Phys. Lett. **103** (20), 204105 (2013).
26. T. Latychevskaia and H.-W. Fink, Opt. Express **21** (6), 7726 (2013).